# Does Criticisms Overcome the Praises of Journal Impact Factor?

Masood Fooladi[1], Hadi Salehi[2], Melor Md Yunus[3], Maryam Farhadi[1], Arezoo Aghaei Chadegani[1], Hadi Farhadi[4] & Nader Ale Ebrahim[5]

[1] Department of Accounting, Mobarakeh Branch, Islamic Azad University, Mobarakeh, Isfahan, Iran

[2] Faculty of Literature and Humanities, Najafabad Branch, Islamic Azad University, Najafabad, Isfahan, Iran

[3] Faculty of Education, Universiti Kebangsaan Malaysia (UKM), Malaysia

[4] School of Psychology and Human Development, Faculty of Social Sciences and Humanities, Universiti Kebangsaan Malaysia (UKM), Malaysia

[5] Research Support Unit, Centre of Research Services, Institute of Research Management and Monitoring (IPPP), University of Malaya, Malaysia

Correspondence: Masood Fooladi, Department of Accounting, Mobarakeh Branch, Islamic Azad University, Mobarakeh, Isfahan, Iran. E-mail: foladim57@gmail.com

**Abstract**

Journal impact factor (IF) as a gauge of influence and impact of a particular journal comparing with other journals in the same area of research, reports the mean number of citations to the published articles in particular journal. Although, IF attracts more attention and being used more frequently than other measures, it has been subjected to criticisms, which overcome the advantages of IF. Critically, extensive use of IF may result in destroying editorial and researchers' behaviour, which could compromise the quality of scientific articles. Therefore, it is the time of the timeliness and importance of a new invention of journal ranking techniques beyond the journal impact factor.

**1. Introduction**

Citation is the reference text and recognition of article information. Generally, each article has a reference section at the end including all references cited in that article. Each reference is called a citation. The frequency of citations to an especial article by other articles means citation count. In addition, citation index as a type of bibliographic database traces the references in a published article. Citation index shows that how many times an article has been cited with other articles. In other words, one can find that which later article cites which earlier articles. Citations are applied to measure the importance of information contained in an article and the effect of a journal in related area of research activity and publications.

The first citation index for published articles in scientific journals is Science Citation Index (SCI), which is founded by Eugene Garfield's Institute for Scientific Information (ISI, previously known as Eugene Garfield Associates Inc.) in 1960 (Nigam & Nigam, 2012.). Then, it was developed to produce the Arts and Humanities Citation Index (AHCI) and the Social Sciences Citation Index (SSCI). The SSCI is one of the first databases developed on the Dialog system in 1972. Small (1973) develops his efforts on Co-Citation analysis as a self-organizing classification mechanism namely "Research Reviews". Garner, Lunin, and Baker (1967) describes that the worldwide citation network has graphical nature. Giles, Bollacker, and Lawrence (1998) introduce the autonomous citation indexing, which can provide the automatic algorithmic extraction and classify the citations for any digital scientific and academic articles. Subsequently, several citation indexing services are created to automate citation indexing including Google Scholar, EBSCOhost, Institute for Scientific Information (ISI) and Elsevier (Scopus).

Among all citation indexing services, ISI, has dominated the citation indexing career. ISI is a part of Thomson Reuters, which publishes the citation indexes in form of compact discs and print format. One can access to the products of ISI through the website with the name 'Web of Science' (WOS). It is possible to access seven databases in WOS including: Social Sciences Citation Index (SSCI), Index Chemicus, Science Citation Index (SCI), Arts & Humanities Citation Index (A&HCI), Current Chemical Reactions, Conference Proceedings Citation Index: Science and Conference Proceedings Citation Index: Social Science and Humanities.

## 2. Impact Factor

One of the most important product of ISI is the journal impact factor (IF), which is first designed by Eugene Garfield. IF is published by ISI as a standardized measure to reveal the mean number of citations to the published articles in particular journal. Journal Citation Reports (JCR) defines the IF as the frequency of citation to the "average papers" published in a journal in an especial year. In effect, IF is almost used by research institutions, universities and governmental agencies to measure the influence and impact of a professional journal comparing with other journals in the same area of research (Kuo & Rupe, 2007). IF is also applied to evaluate the individual profession of scientists and researchers. Therefore, editors have a high tendency to increase the IF of their journals.

In order to calculate the IF of a journal, the number of citations to the published papers in a journal, which is reported in the JCR year, is divided by the total number of papers published in that journal during two previous years. Citation may be comes from articles published in the same journal or different proceedings, journals or books indexed by WOS. As an example, an IF of 3.0 reveals that, on average, the papers published in a particular journal within two years ago, have been cited three times. Although the IF calculation for journals is based on articles published and cited in the previous two years, one can calculate the average citation using longer time periods.

The 5-year journal IF is the number of citations to a particular journal divided by the total number of papers published in the journal since five years ago. In certain fields of study, the 5-year journal IF is more appropriate than 2-years because the body of citations in 5-years basis might be large enough to make a reasonable comparison. In some area of research, it takes a long time, more than two years, to publish a work and response to previous articles.

The aggregate IF is applied for a subject category rather than a journal only (Khan & Hegde, 2009). The calculation is the same way as the IF for a journal, but it considers the total number of citations to all journals in a particular subject group and the number of papers published in these cited journals. The median IF refers to the median value of IF for all journals in the subject category. The number of citations to a journal divided by the total papers published in the journal in current year is known as immediacy index. The journal's half-life in each year refers to the median age of the article cited in JCR. As an example, if the value of journal's half-life equals to six in 2007, it means that the citations from 2002 until 2007 to the journal are 50% of all the citations to the journal in 2007. The journal ranking table displays the ranking of a journal in related subject category based on the journal IF (Khan & Hegde, 2009).

## 3. Praises for Impact Factor

IF as a measure of journals' influence has several advantages and benefits for researchers as well as librarians, knowledge managers and information professionals. Some of the advantages are listed below:

1) The primary advantage of impact factor is that it is very easy to measure.

2) Another advantage of IF is to mitigate the absolute citation frequencies. Since large journals provide a considerable body of citable literature which cites to the small journals, IF discounts the benefits of large journals in favour of small journals. In addition, there is a tendency to discount the benefits of high frequently published journals in favour of less frequently published journals and also matured journals over the newly established journals. In other words, journal IF is a remarkable measure for evaluating the journals since it offsets the benefits of age and size among the journals.

3) IF is a useful tool to compare different research groups and journals. It is commonly applied to administer scientific library collections. Librarians, knowledge managers and Information professionals encounter with limited budgets when they aim to select journals for their institutions and departments. ISI provide them with the IF as a measure to choose most frequently cited journals (Khan & Hegde, 2009). About 9000 social science and science journals from 60 countries are indexed by Web of Knowledge.

4) IF for indexed journals are greatly available and easy to understand and use. IF is more acceptable and popular than other alternative measures (Khan & Hegde, 2009).

## 4. Criticisms for Impact Factor

However, IF has several inherent shortcomings overcoming the advantages of IF. The following problems refer to the calculation method of IF:

1) It should be noted that Thomson Scientific is a commercial institution, which sells its products and evaluations to research institutions and publishers. Data and measures of IF are not publicly established and it is not easy for scientists to access these data and measures. Therefore, it is not subject to a peer review process. Researchers demand a fully transparency over how Thomson Scientific collect data and calculates citation metrics. Thomson Scientific is unable to clarify the used data to support its published IF and hence its IF is not reliable. Generally, findings of a scientific article could not be accepted by scientists when the primary data is not available.

2) The coverage of database is not complete since Thomson Scientific excludes some specific types of sources from the denominator. As an example, books (as a source for citations) are excluded from the database. Citations from journals which are not indexed in the ISI are not considered in IF calculation (Khan & Hegde, 2009). Falagas, Kouranos, Arencibia-Jorge, and Karageorgopoulos (2008) assert that one of the major shortcomings of journal IF is considering only "citable" papers mainly original papers and reviews.

3) IF could not be an appropriate measure because it is an arithmetic mean of number of citations to each paper and the arithmetic mean is not an appropriate measure (Joint Committee on Quantitative Assessment of Research 2008). IF does not consider the actual quality of research articles, their impressiveness, or the long-term impact of journals. It should be noted that IF and citation indexes indicate a specific, and relatively uncertain, type of performance. They do not provide a direct measure of quality (Moed, 2005). Therefore, Using the IF to derive the quality of individual articles or their authors seems to be idle (Baum, 2011). In addition, the validity and appropriate use of the IF as a measure of journal importance is a subject to controversy and another independent examination might reproduce a different IF (Rossner, Van Epps, & Hill, 2007). IF make more controversy when it is used to assess the articles published in the journals. IF appraises the reputation of the publishing journal not the quality of the content in the single papers. In addition, Falagas et al. (2008) assert that IF takes into account received citations only in a quantitative manner. It is conceivable that the published papers in a journal have a greater influence on science if these papers are cited by articles with a higher scientific quality.

4) The two-year or five-year window for measuring the IF might be logical for some fields of studies with a fast moving process of research, while it is not reasonable for some fields of study, which requires a long period for research or empirical validation and also takes a long time for review. These kinds of research might take a time longer than two years to be completed and then published. Therefore, citation to the original papers will not be considered in the IF of publishing journal. In addition, articles, which are published in many years ago and are still cited, have a significant impact on research area, but unfortunately, citation to these articles will not be considered due to their old age.

5) Distributing the same value of IF to each article published in a same journal leads to excessive variability in article citation, and provides the majority of journals and articles with the opportunity to free ride on a few number of highly cited journals and articles. Only some articles are cited anywhere near the mean journal IF. In effect, IF highly overstates the impact of most articles, while the impact of rest articles is greatly understated because of the few highly cited articles in the journal (Baum, 2011). Therefore, journal IF could not portray the individual impact of single articles and it is not much related to the citedness of individual articles in publishing journal (Yu & Wang, 2007). Seglan (1997) states that IF is wrongly used to estimate the importance of a single article based on where it is published. Since IF averages over all published papers, it underestimates the importance of the highly cited papers whereas overestimating the citations to the average publications.

6) Although, it is feasible to evaluate the IF of the journals in which an individual has published papers, the IF is not applicable to individual scientists or articles. ISI does not distribute a JCR for the humanities. The relevant number of citations received by a single paper is a better measure of importance and influence of individual papers as well as its authors (Khan & Hegde, 2009). Garfield (1998) caution regarding the "misuse in evaluating individuals" by IF due to a wide variation from paper to paper within an individual journal.

7) The IF is highly depended on different disciplines such as physical sciences and mathematical because of the speed with which articles are cited in a research area (Van Nierop, 2009). It is not logical to use IF as a measure to compare journals across different disciplines. The absolute value of IF does not make any sense. For example, a journal with the value of IF equal to two could not have much impact in fields and disciplines such as Microbiology, while it would be impressive in Oceanography. Therefore, IF as a measure of comparison between different fields and disciplines could be considered as invalid. In addition, this comparison has been widely made not only between the journals, but also between scientists or university departments, while, absolute IF could not estimate the scientific level of department. In an evaluation program, especially for doctoral degree granting institutions, reviewers consider the IF of journals in examining the scholarly outputs. It makes a bias in which some types of researches are undervalued and falsifies the total contribution of each scientist (Khan & Hegde, 2009).

8) IF is not relevant in some disciplines such as engineering, where the main scientific outputs are technical reports, conference proceedings and patents. Regarding the literature, IF is not a relevant measure because books constitute the most important publications in literature, which cite other books.

9) The number of citations from journals in less universal languages and less-developed countries is understated, since only ISI database are utilized to define the IF.

It is argued that the simple methodology applied in the calculation of IF provides editors with the opportunity to use different practices in order to inflate the impact factor of their journals (Garfield, 1996; Hemmingsson, Skjennald, & Edgren, 2002; The PLoS Medicine Editors, 2006). Reviewers may request from the authors to expand the article's citation before publishing the articles. This is usual in peer review process in order to improve the quality of article. On the other hand, IF indicates the importance of a journal in related subject category. Therefore, editors have a high tendency to increase their journal IF (Yu, Yang, & He, 2011). There are several methods for a journal to increase its IF. One policy is to manipulate the IF (Dong, Loh, & Mondry, 2005). References to a journal can be managed in order to increase the number of citations to a journal and hence raise the IF (Smith, 1997). Therefore, these journals seem to have additional impact and their quality and quantity will be increased. However, there is no actual improvement in the journal impact and influence. This manipulation of IF will only result in artificial evaluation.

1) Self-citation is a common way to manipulate and increase the IF of journals (Miguel & Marti-Bonmati, 2002; Fassoulaki, Papilas, Paraskeva, & Patris, 2002; Falagas & Kavvadia, 2006; Falagas & Alexiou, 2007; Yu et al., 2011). Editorial board may force authors to expand the article's citations not necessarily to improve the quality of scientific articles but they aim to inflate the IF of their journal in order to artificially raising the journal's scientific reputation (Wilhite & Fong, 2012). This is a business policy, which is called coercive citation. In effect, editors require authors to cite papers published in the same journal (self-citation) before the journal accepts to publish the article even if these citations are not relevant to the research. In their study, (Wilhite & Fong, 2012) find that 20% of researchers in psychology, sociology, economics and multiple business disciplines has experienced coercive citation. They find that journals with a lower IF have a high tendency for coercive citation in order to inflate their IF. Another study documents that coercive citation is common in other scientific disciplines.

2) Another policy to manipulate the IF is to focus on publishing review articles rather than research articles because review articles have a large body of literature, which are highly citable (Andersen, Belmont, & Cho, 2006; Falagas et al., 2008; Arnold & Fowler, 2011). Therefore, review articles provide the higher IF for publishing journals and these journals will have the highest IF in their respective research area. Journals have a tendency to publish a large number of their articles, which are likely to receive a high citation in near future or deny accepting papers such as case report in medical journals, which are not expected to be cited.

Considering the above drawbacks of IF, it is suggested that IF as a measure of journal evaluation should have certain characteristics (Offutt, 2008). It should be independent from the length and number of the papers published in the journal. In addition, it should consider essential advances in the area over short term or incremental contributions. IF should have relative stability in a year to year basis. The most important is that IF should evaluate the impact of a journal during a long period since some articles are still cited after 10, 20, or even 50 years from the date of publication. In addition, we should consider the warnings by Garfield against using to compare different IF and scientific field. Beside the IF we can consider other internet databases to calculate the IF of journals or articles which is collaborating with Elsevier.

The journal IF is an indicator of journal performance, and as such many of the suggestions for performance indicators could equally be used to the application of journal IF (Greenwood, 2007). In November 2007 the European Association of Science Editors (EASE) published an official statement suggesting that journal IF can be applied only and cautiously for assessing and comparing the importance and impact of journals, but not for evaluating the single articles, and of course not for appraising the research programmes or researchers (EASE, 2007). The International Council for Science (ICSU) Committee on Freedom and Responsibility in the conduct of Science (CFRS) issued a "Statement on publication practices and indices and the role of peer review in research assessment" in July 2008, which suggest to consider only a limited number of articles published per year by each scientist or even give a penalty to scientists due to excessive number of publications in each year (Smith, 1997). The Deutsche Forschungsgemeinschaft (2010), German Foundation for Science, issued new strategies to assess only papers and no bibliometric information on candidates in decision making process

regarding:

> "...performance-based funding allocations, postdoctoral qualifications, appointments, or reviewing funding proposals, [where] increasing importance has been given to numerical indicators such as the h-index and the impact factor".

The citation style of manipulated journals, especially the number of self-citations in the last two year appears to be unusual. It can be seen that the self-cited rate of manipulated journals is higher than that of the normal journal. Therefore, self-citation rate can be used as a measure to specify the manipulated journals and calculate the real IF by deducting the self-citations from the reported IF. However, it is a difficult task to specify manually each journal in the JCR database. As discussed, journal might publish a larger number of review articles in order to inflate their IF. Therefore, the Thomson Scientific website can provide directions for removing review articles from the calculation (Khan & Hegde, 2009).

## 5. Conclusion

Although IF is a popular measure of quality for journals, we conclude that journal IF has its own restrictions. We believe that there should be more examination of whether and how IF evaluates journal quality before it is broadly accepted as a measure of journal quality. The IF measurement would certainly never assess the peer review process of journals (Offutt 2008). It means that IF is not an appropriate measure to assess the quality of published articles. One important factor in assessing a progress case is to consider the number of published articles but it is an imperfect method because it considers the number of articles only instead of the quality of publications. However, it is very hard to assign professionals to read articles and give an independent opinion about the impact of particular scientist on his or her research area.

The implication for researchers and journals is that they should not rely only on this indicator. If we do not consider the above limitations associated with IF, decisions made based on this measure is potentially misleading. Generally, a measure will fall into disuse and disrepute among the scientific community members, if it is found to be invalid, unreliable, or ill-conceived. Although IF has many limitations as discussed in this paper, it does not lose its reputation and application by the scientific society. Indeed, IF attracts more attention and being used more frequently by scientists and librarians, knowledge managers and information professionals. Critically, extensive use of IF may result in destroying editorial and researchers' behaviour, which could compromise the quality of scientific articles. Calculation of IF and policies to increase the IF by journals, may push researchers to consider publication as a business rather than contribution to the area of research. It is not fare that we should rely on such a non-scientific method which IF appraises the quality of our efforts. It is the time of the timeliness and importance of a new invention of journal ranking techniques beyond the journal impact factor. There should be a new research trend with the aim of developing journal rankings that consider not only the raw number of citations received by published papers, but also the influence or importance of documents which issue these citations (Palacios-Huerta & Volij, 2004; Bollen, Rodriguez, & van de Sompel, 2006; Bergstrom, 2007; Ma, Guan, & Zhao, 2008). The new measure should represent scientific impact as a function not of just the quantity of citations received but of a combination of the quality and the quantity.